\documentclass{icrc2009}

\usepackage{graphicx}   
\usepackage[caption=false]{caption}    
\usepackage[font=footnotesize]{subfig} 
\usepackage{fixltx2e}
\usepackage{url}
\usepackage{textcomp}
\usepackage{amsfonts}
\usepackage{amssymb}

\newcommand{\shorttitle}[1]%
{\markboth{Proceedings of the 31\MakeLowercase{$^{st}$} ICRC, {\L}\'{o}d\'{z} 2009}{#1} }
\newcommand{\etal}{\MakeLowercase{\textit{et al. }}} 
\newcommand{\td}{\textdegree}
\newcommand{\be}{\begin{enumerate}}
\newcommand{\ee}{\end{enumerate}}

\begin{document}
\title{Observations of Supernova Remnants with VERITAS}

\author{\IEEEauthorblockN{Brian Humensky\IEEEauthorrefmark{1} for the VERITAS Collaboration\IEEEauthorrefmark{2}}
                            \\
\IEEEauthorblockA{\IEEEauthorrefmark{1}University of Chicago Enrico Fermi Institute, Chicago, IL 60637, USA (humensky@uchicago.edu)}
\IEEEauthorblockA{\IEEEauthorrefmark{2}see R.A. Ong et al (these proceedings) or http://veritas.sao.arizona.edu/conferences/authors?icrc2009}
}

\shorttitle{T. B. Humensky \etal Observations of Supernova Remnants with VERITAS}
\maketitle

\begin{abstract}
The study of shell-type supernova remnants is a key science focus for the VERITAS TeV telescope array. Supernova remnants (SNRs) are widely considered to be the strongest candidate for the source of cosmic rays below the knee around $10^{15}\ \textrm{eV}$. This presentation will highlight new VERITAS results including new measurements of the spectra of Cas A and IC 443. These results and their implications for the nature of the cosmic rays - hadronic or electronic - accelerated in the remnants will be discussed. \\
  \end{abstract}

 
\section{Introduction}
Fermi acceleration in supernova remnants (SNRs) has long been considered as one of the likely explanations for the origin of cosmic rays (CRs) up to the knee at $3\cdot 10^{15}\ \textrm{eV}$. This scenario drives the observations of SNRs in the very-high-energy (VHE, $E>100\ \textrm{GeV}$) gamma-ray band.  Measurements of the VHE morphology and energy spectrum, when combined with multiwavelength observations, hold promise to constrain models of diffusive shock acceleration and the potential to discriminate between gamma-ray production by leptonic and hadronic cosmic rays.  For example, the deep H.E.S.S. observations of RX J1713.7-3946 reveal an energy spectrum extending beyond $30\ \textrm{TeV}$, implying particle acceleration to well beyond $100\ \textrm{TeV}$ \cite{Aharonian2007a}. Unfortunately, in many cases (including RX J1713.7-3946) uncertainty in the environment (age and distance of remnant, density of interstellar medium) makes it difficult to discriminate conclusively between hadronic and leptonic scenarios for the gamma-ray emission.

This paper discusses detections of VHE gamma rays from the supernova remnants Cassiopeia A and IC 443. An upper limit on the ``Forbidden Velocity Wing'' FVW 190.2+1.1, a possible supernova remnant, is discussed in \cite{Holder2009a}.  A summary of the full VERITAS observation program on supernova remnants and pulsar wind nebulae is given in \cite{Humensky2009b} in these proceedings.

\section{Observations and Analysis}
VERITAS \cite{Holder2006a} consists of four $12\textrm{-m}$ telescopes located at an altitude of $1268\ \textrm{m}$ a.s.l. at the Fred Lawrence Whipple Observatory in southern Arizona, USA (31\td\ 40' 30'' N, 110\td\ 57' 07'' W). Each telescope is equipped with a 499-pixel camera of $3.5\textrm{\td}$ field of view.  The array, completed in the spring of 2007, is sensitive to a point source of 1\% of the steady Crab Nebula flux above $300\ \textrm{GeV}$ at $5 \sigma$ in less than 50 hours at 20\td\ zenith angle.

Observations are taken in wobble mode \cite{Fomin1994a}, with the telescope pointing offset by 0.5\textdegree-0.7\td\ from the target direction for roughly equal amounts of time in each of four equally spaced directions.  

The data analysis is described more fully in \cite{Daniel2007a}.  In brief, pixel gains are flatfielded using laser flashes recorded during nightly calibration runs. A two-pass cleaning algorithm with thresholds of 5 and 2.5 times the pedestal RMS is used; isolated pixels which survive the image cleaning are then eliminated.  The images are parameterized according to their first and second moments \cite{Hillas1985a}, and then the shower direction and core location are determined.  Lookup tables provide expected values for the image width, length, and energy, and are used to construct the gamma-hadron separation parameters Mean Scaled Width (MSW) and Length (MSL), and an estimate of the shower energy \cite{Konopelko1995a}.  Images are included in the reconstruction if they have at least five pixels, if the centroid is less than 1.43\textdegree\ from the center of the camera, and if the image size (sum of the pixel charges) is at least 75 p.e.  At least two good images per event are required.  Events in which only the two telescopes sharing the shortest baseline have good images are rejected, as their $35\textrm{-m}$ separation is too small for good stereo reconstruction.  Cuts on MSW and MSL have been optimized for weak sources with a Crab-like spectrum.  The background is estimated using the ``ring background model'' \cite{Aharonian2005a}.  Significances are calculated according to equation 17 of \cite{Li1983a}.  All results are verified with an independent analysis.

\section{Cassiopeia A}
Cassiopeia A (Cas A) is a young remnant, believed to be about 300 years old and at a distance of $3.4\ \textrm{kpc}$.  It is the brightest and among the best-studied radio sources in the sky \cite{Kassim1995a}.  The synchrotron radiation from Cas A extends from radio wavelengths to hard X-rays \cite{Allen1997a,Favata1997a}.  It has a diameter of 5 arcmin, making it a point source for TeV telescopes.  Cas A was not detected by EGRET, and is not seen in the Fermi Gamma-ray Space Telescope (FGST) bright-source list \cite{Abdo2009a}.  Cas A was first detected in TeV gamma rays by HEGRA \cite{Aharonian2001a}, with a flux of $\sim 3.3\%$ Crab above $1\ \textrm{TeV}$.  The detection was confirmed by MAGIC \cite{Albert2007b} at $5.3\ \sigma$ in a 47-hour observation, yielding a flux above $1\ \textrm{TeV}$ of $(7.3 \pm 0.7_{stat} \pm 2.2_{sys}) \times 10^{-13}\ \textrm{cm}^{-2}\textrm{s}^{-1}$ and a spectral index of $2.3 \pm 0.2_{stat} \pm 0.2_{sys}$, consistent with the HEGRA results.  Cas A is an interesting TeV SNR because of its low age, detailed observations across all wavelengths, and relatively simple local environment, making it attractive for constraining models of particle acceleration in shocks.

Cas A was observed in October/November, 2007 with four telescopes. After data quality selection (for rate stability and good weather) and correction for deadtime, an exposure of 20.3 hours remains with an average zenith angle of 30\textdegree.  

\begin{figure}[!t]
  \centering
  \includegraphics[bb=0 35 567 480,clip,width=2.5in]{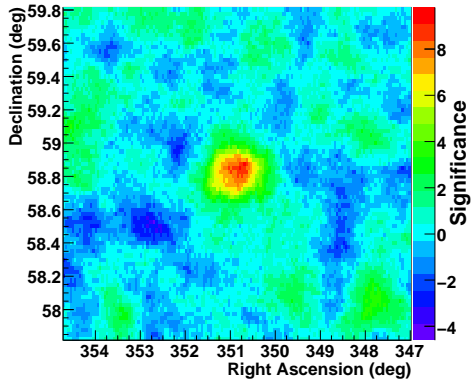}
  \caption{Cas A significance map, showing a strong point-like signal at the location of Cas A.}
  \label{fig1}
 \end{figure}

\begin{figure}[!t]
  \centering
  \includegraphics[bb=0 35 567 265,clip,width=2.5in]{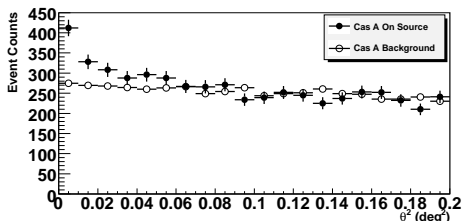}
  \caption{Cas A $\theta^2$ distribution, indicating the number of events as a function of the square of the separation angle from the center of Cas A.  The filled circles are for the on-source region and the empty circles are an estimate of the background via the reflected-region method.}
  \label{fig2}
 \end{figure}

Preliminary results, previously show in \cite{Humensky2009a}, are given here, and final results will be presented at the ICRC.  Figure~\ref{fig1} shows the significance map for the Cas A field and Figure~\ref{fig2} shows the distribution of events as a function of the square of the angular distance from the source location.  Clear peaks are seen at the location of Cas A, with a significance of $9.8\ \sigma$ and a flux (determined by comparing to the rate of gamma rays from the Crab Nebula) of $\sim 3\ \%$ Crab, consistent with the HEGRA and MAGIC results.  The emission is consistent with a point source within the instrument's angular resolution, and the spectral analysis is in progress.  

\section{IC 443}
IC 443 is one of the classic examples of a SNR interacting with a molecular cloud, and this interaction greatly affects its evolution.  VHE emission coincident with the site of the shock/cloud interaction arguably provides a smoking gun for the acceleration of hadronic cosmic rays.  IC 443 is at a distance of $1.5\ \textrm{kpc}$  and has a diameter of $\sim 0.75\textrm{\textdegree}$.  The northeast hemisphere is interacting with an H\footnotesize II\normalsize\ region, and the southwest hemisphere is interacting with a molecular cloud at several points.  The cloud has a mass of $\sim 10^4\ \textrm{M}_{\odot}$ \cite{Butt2003a}, mostly along the line of site to IC 443; maser emission has been observed in this region \cite{Claussen1997a}.  The age of IC 443 is unclear; various estimates place it in the range 3-30 kyr (see, e.g., \cite{Troja2008a,Lee2008a}).  The PWN CXOU J061705.3+222127 is located at the southern edge of the remnant \cite{Olbert2001a,Bocchino2001a,Gaensler2006a}.  Both MAGIC and VERITAS have reported TeV emission from the central region of the remnant, coincident with the maser emission \cite{Albert2007a,Humensky2007a}.  An EGRET GeV source also overlaps the remnant \cite{Hartman1999a}, and recent detections by AGILE \cite{Pittori2009a} and FGST \cite{Abdo2009a} have been reported.

IC 443 was observed by VERITAS during two periods: with three telescopes in February/March, 2007 (during the commissioning phase of the array) for 20 hours; the PWN was the target for these observations.  A further 25 hours of observations were taken in October/November, 2007 using the complete four-telescope array and targeting the location of the emission observed previously, 06 16.9 +22 33.  After quality selection and deadtime correction, the data set has a livetime of 37.1 hours and an average zenith angle of 18\textdegree.

One systematic issue particular to this data set is the presence of two bright stars in the IC 443 field: Eta Gem (V band magnitude 3.31), located 0.53\td\ from the PWN; and Mu Gem (V band magnitude 2.87), located 1.37\td\ from the PWN. The high flux of optical photons from the stars requires that several PMTs in each camera be turned off during the observations and produces increased noise levels in the signals of nearby PMTs.  These effects reduce the exposure in the vicinity of the stars and degrade the angular resolution.  To counteract these effects while studying the morphology of the source, a high telescope-multiplicity requirement, described below, was selected.

\begin{figure}[!t]
  \centering
  \includegraphics[bb=0 35 567 480,clip,width=2.5in]{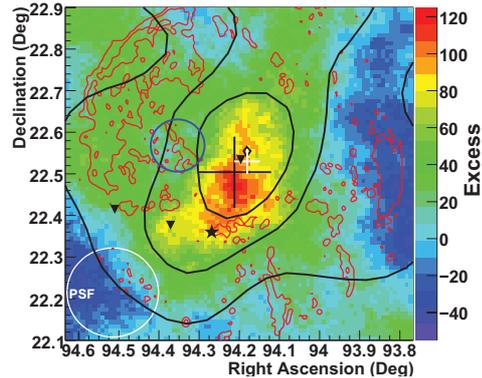}
  \caption{Inner $0.8\textrm{\td}$ of the acceptance-corrected excess map for the IC 443 field. The black cross-hair indicates the centroid position and its uncertainty (statistical and systematic added in quadrature), and the white cross-hair likewise indicates the position and uncertainty of MAGIC J0616+225 \cite{Albert2007a}.  Red contours: optical intensity \cite{McLean2000a}.  Thick black contours: CO survey \cite{Huang1986a}; black star: PWN CXOU J061705.3+222127 \cite{Olbert2001a}; open blue circle: 95\% confidence radius of 0FGL J0617.4+2234; and filled black triangles: locations of OH maser emission (\cite{Claussen1997a}, \cite{Hewitt2006a}, J. W. Hewitt, private communication). The white circle indicates the PSF of the VERITAS array.}
  \label{fig3}
 \end{figure}

\begin{figure}[!t]
  \centering
  \includegraphics[bb=0 35 567 425,clip,width=2.5in]{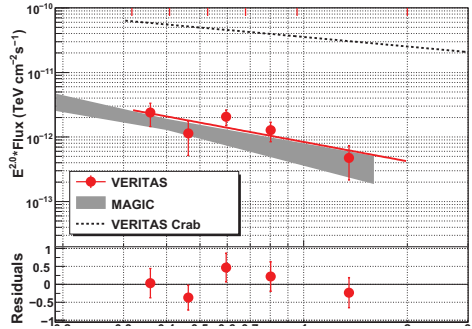}
  \caption{Spectrum of IC 443, scaled by $E^2$.  Red points are the VERITAS spectrum (red tick marks along top indicate bin edges).  The red line is a power-law fit (see text), with residuals to the fit plotted in the lower box.  The MAGIC spectrum is indicated by the gray band and extends down to $90\ \textrm{GeV}$.  VERITAS error bars and MAGIC error band reflect statistical errors only.  The Crab spectrum (V. Acciari et al. 2009, in preparation) is shown as a dashed line for comparison.}
  \label{fig4}
 \end{figure}

A standard point-source analysis of the full data set yields a maximum significance of 8.3 standard deviations ($\sigma$) before trials (7.5 $\sigma$ post trials, accounting for a blind search over the region enclosed by the shell of IC 443) at $06^{\textrm{h}}16^{\textrm{m}}49^{\textrm{s}}\textrm{+22\td} 28'30''\ (\textrm{J2000})$ and a significance of 6.8 $\sigma$ at the location of MAGIC J0616+225 (see \cite{Acciari2009a} for full details).  Table~\ref{tbl-1} summarizes the results of this analysis at the location of maximum significance, listing the counts falling within the point-source integration radius of 0.112\td\ ($on$), the counts integrated in a background ring spanning radii 0.6-0.8\td\ ($off$), the ratio of $on$ exposure to $off$ exposure ($\alpha$), and the resulting number of excess counts and significance.  Figure~\ref{fig3} shows the inner $0.8\textrm{\td}$ of the excess map.  To study the source morphology, the best-reconstructed gamma rays are selected by requiring that an event must have images from all participating telescopes survive the pre-selection cuts.  This requirement increases the analysis energy threshold by $\sim$15\% and reduces the excess by $\sim$40\%.  The centroid and intrinsic extension of the excess are characterized by fitting an azimuthally symmetric two-dimensional Gaussian, convolved with the PSF\footnote{The PSF is characterized as a sum of two, two-dimensional Gaussians describing a narrow core and a broader tail.  It is determined from data taken on the Crab Nebula, which is a point source at these energies \cite{Albert2008a}.} of the instrument, to an acceptance-corrected uncorrelated excess map with a bin size of 0.05\td.  The PSF has a 68\% containment radius of 0.11\td.  The centroid determined in this way is located at $06^{\textrm{h}}16^{\textrm{m}}51^{\textrm{s}}\textrm{+22\td} 30'11''\ (\textrm{J2000})\ \pm 0.03\textrm{\td}_{stat} \pm 0.08\textrm{\td}_{sys}$, consistent with the MAGIC position, and the extension is $0.16\textrm{\td} \pm 0.03\textrm{\td}_{stat} \pm 0.04\textrm{\td}_{sys}$.  Due to the effects of the nearby bright star and the non-Gaussian nature of the excess, a large systematic error is currently assigned to the centroid.  The difference in extension between this work and the point-like source detected by \cite{Albert2007a} can be explained by the difference in angular resolution and sensitivity between VERITAS and MAGIC, the latter of which has an angular resolution of $\sim$$0.15\textrm{\td}$ for 68\% containment and a sensitivity above $200\ \textrm{GeV}$ of 2\% of the Crab Nebula flux in 50 hours \cite{Albert2008a}.  The photon spectrum, integrated within a radius of 0.235\td, is shown in Figure~\ref{fig4}.  The photon spectrum is well fit ($\chi^2/\textrm{ndf} = 3.1/3$) by a power law $dN/dE = N_0 \times (E/\textrm{TeV})^{-\Gamma}$ with a normalization of $(8.38 \pm 2.10_{stat} \pm 2.50_{sys}) \times 10^{-13}\ \textrm{TeV}^{-1}\ \textrm{cm}^{-2\ }\textrm{s}^{-1}$ and an index of $2.99 \pm 0.38_{stat} \pm 0.30_{sys}$.  The integral flux above a threshold energy of $300\ \textrm{GeV}$ is $(4.63 \pm 0.90_{stat} \pm 0.93_{sys}) \times 10^{-12}\ \textrm{cm}^{-2}\ \textrm{s}^{-1} $ (3.2\% of the Crab Nebula flux).

\begin{table}[!h]
\caption{Analysis results for VERITAS observations of IC 443.\label{tbl-1}}
\begin{tabular}{lccccc}
\hline\hline
 & $on$ & $off$ & $\alpha$ & excess & significance \\
      &      &       &          &        & ($\sigma$)   \\
\hline
data set 1 & 393 & 5152 & 0.0610 & 78.7 & 4.1  \\
data set 2 & 609 & 7331 & 0.0601 & 168.1 & 7.3  \\
combined   & 1002 & 12483 & 0.0604 & 247.5 & 8.3 \\
\hline
\end{tabular}
\end{table}

The TeV centroid is $\sim 0.15\textrm{\td}$ from the position of the PWN and $\sim 0.03\textrm{\td}$ from the nearby, bright maser emission coincident with clump G of \cite{Huang1986a}.  Weaker regions of maser emission lie along the southern rim of the remnant, east of the PWN \cite{Hewitt2006a}.  The VHE emission coincides with the densest region of the molecular cloud.  While the Fermi source 0FGL J0617.4+2234 is displaced from the centroid of the VHE emission by $\sim$0.15\td, this is consistent at the 95\% level with the combined errors quoted by Fermi and VERITAS.

The VHE gamma-ray emission observed by VERITAS and MAGIC is offset from the location of the PWN by $\sim$$10\textrm{-}20$ arcmin, similar to other offset PWNe such as HESS J1825-137 \cite{Aharonian2006a}. The emission is consistent with a scenario in which the VHE emission arises from inverse Compton scattering off electrons accelerated early in the PWN's life, as in \cite{Bartko2008a}.  If the VHE emission is indeed associated with IC 443 at a distance of $1.5\ \textrm{kpc}$, the luminosity in the energy band $0.3\textrm{-}2.0\ \textrm{TeV}$ is $4 \times 10^{32}\ \textrm{erg s}^{-1}$.  The spin-down luminosity of the pulsar has been estimated as $\sim$$10^{36}\ \textrm{erg s}^{-1}$ \cite{Olbert2001a,Bocchino2001a} and $5 \times 10^{37}\ \textrm{erg s}^{-1}$ \cite{Gaensler2006a}.  If the PWN association is correct, the VHE luminosity in the $0.3\textrm{-}2.0\ \textrm{TeV}$ band is less than $\sim$$0.04\textrm{\%}$ of the spin-down luminosity, at the lower end of the $\sim$$0.01\textrm{-}10\%$ range of observed VHE efficiencies for other PWNe \cite{Gallant2007a}.  

Alternatively, Figure~\ref{fig3} can be interpreted as hadronic cosmic-ray acceleration and subsequent interaction with the molecular cloud, which would provide a high density of target material for the production of VHE gamma rays.  The steep VHE spectrum can be explained either as an energy-dependent rate of diffusion of cosmic rays out of the cloud \cite{Aharonian1996a}, or a low maximum energy to which particles were accelerated prior to the shock hitting the cloud.    

Further GeV and TeV gamma-ray observations will aid in elucidating the nature of the particle acceleration associated with IC 443.  The $\sim$0.15\td\ separation between the centroids of 0FGL J0617.4+2234 and VER J0616.9+2230, though not statistically significant, may hint at an energy-dependent morphology.  This will be explored with future observations by VERITAS, and has the potential to discriminate between the leptonic and hadronic scenarios sketched above.  The relocation of one of the telescopes during the summer of 2009 as part of the VERITAS upgrade program \cite{Otte2009a} will improve both the angular resolution and sensitivity of the array, assisting with these studies.  Whether or not 0FGL J0617.4+2234 and VER J0616.9+2230 are directly related, a break in the energy spectrum between the GeV and TeV bands is required.  Future observations by VERITAS will improve the statistics above $100\ \textrm{GeV}$, anchoring the measurement of the VHE cut-off to the gamma-ray emission.  The combined spectrum from $100\ \textrm{MeV}$ to several TeV will provide powerful constraints on the emission mechanism(s)~\cite{Gaisser1998a}.

\section{Conclusions}
VERITAS has made strong detections of both the Cassiopeia A and IC 443 supernova remnants.  Cas A is observed to have a flux of $\sim3\%$ Crab and appears pointlike within the instrument's resolution.  These results are consistent with previous observations, and final results, including a spectrum and an upper limit on the source extension, are in preparation.  

The emission from IC 443 is extended, yielding a sigma of $\sim 0.16\textrm{\textdegree}$ when fit with a two-dimensional gaussian.  It has an integral flux of $3.2\%$ Crab above $300\ \textrm{GeV}$.  The TeV emission is coincident with both the densest part of the molecular cloud that IC 443 is interacting with and the maser emission observed within that cloud, strongly suggesting that the gamma rays are produced by the interaction of hadronic cosmic rays with cloud material.  An alternative explanation in terms of an association with the PWN CXOU J061705.3+222127 is also possible.


The VERITAS survey of the Cygnus region of the Galactic plane, spanning Galactic longitude $67 < l < 82$ and latitude $-1 < b < 4$, covers a number of other supernova remnants such as $\gamma$Cygni and CTB 87 and is discussed elsewhere in these proceedings \cite{Weinstein2009a}.

\section{Acknowledgements}
This research is supported by grants from the US Department of Energy, the US National Science Foundation, and the Smithsonian Institution, by NSERC in Canada, by Science Foundation Ireland, and by STFC in the UK. We acknowledge the excellent work of the technical support staff at the FLWO and the collaborating institutions in the construction and operation of the instrument.

\end{document}